\documentclass[12pt]{article}
\usepackage{amsfonts}
\usepackage{amsmath}
\title{\textbf{the Non-Gaussianity of Racetrack Inflation Models}}
\author{\textbf{Cheng-Yi Sun\footnote{cysun@mails.gucas.ac.cn}
\ and \ De-Hai Zhang\footnote{dhzhang@gucas.ac.cn}}\\
Department of Physics,\\
The Graduate School of The Chinese Academy of Sciences,\\
Beijing 10049, P.R.China.}

\begin{document}
\maketitle
\begin{abstract}
In this paper, we use the result in \cite{a0510709} to calculate the
non-Gaussianity of the racetrack models in \cite{t0406230,t0603129}.
The two models give different non-Gaussianities. Both of them are
reasonable.
\end{abstract}

\ \ \ \ PACS: 98.80.Cq, 98.80.Jk, 04.20.Gz

\ \ \ \ {\bf {Key words: }}{non-Gaussianity, non-linear parameter,
racetrack inflation}

\section{Introduction}

In cosmology, the inflation paradigm plays an important role. The
central idea of inflation is very simple. But it has proved
difficult to discriminate among the large number of different models
that have been developed to date \cite{p9807278}. Both the simplest
classes of inflation models and the most complicated classes may
predict Gaussian-distributed perturbations and nearly
scale-invariant spectra of the primordial density perturbations
\cite{CISL}. However it is believed that the deviation away from the
Gaussian statistics may be a potential powerful discriminant between
the competing inflationary models. On the other hand, physicists are
always trying to understand cosmological inflation within the deeper
theory, e.g. string theory. In the past two years, continued
progress has been made in identifying how the inflation arises from
within string theory. Recently, the racetrack inflation models,
which are based on the Calabi-Yau compactification of type
\textrm{II}B String, attract the attention. Ref. \cite{t0406230}
suggests a simple racetrack inflation model, where only a single
K\"{a}hler modulus is used. And in Ref. \cite{t0603129} a
complicated  racetrack inflation model is given with two K\"{a}hler
moduli (See \cite{t0603129} for comparison between the two models.).
In this paper, we try to calculate the non-Gaussianities, the
non-linear parameter $f_{NL}$, of the two models.

The racetrack models give effectively multi-field inflation models,
and, for a multiple-field inflation model, the expression of
$f_{NL}$ is very complicated. Fortunately, in \cite{a0504045}, Lyth
and Rodriguez have shown that the non-Gaussianity of the curvature
perturbation in multiple field models can be simply expressed in the
so-called ``$\delta N$-formalism" \cite{a0411220}. Further, in
\cite{a0506056}, the authors have given the expression of $f_{NL}$
involving the metric of the field space explicitly. Further more, in
\cite{a0510709}, $f_{NL}$ has been expressed in terms of the
slow-rolling parameters. In this paper, we would adopt the result in
\cite{a0510709} to calculate the non-linear parameter. Here we
should note that the result in \cite{a0510709} is obtained by
neglecting the non-adiabatic perturbations and the intrinsic
non-Gaussianity of the fields (See \cite{a0510709} for details.). Of
course, for the two racetrack models, due to the interacting terms
in the effective actions the non-adiabatic perturbations would be
generated unavoidably. And the intrinsic non-Gaussianity of the
field perturbations exists too. However, the assumption of the
Gaussian-distributed and adiabatic perturbations is in good
agreement with the observation \cite{WMAP1,WMAP3}. So we suppose
that in the racetrack models, this assumption is still good enough.
Then we expect the dominant non-Gaussianity would be obtained by
using the result in \cite{a0510709}.

In this paper, we first give a brief summary of the result in
\cite{a0510709}. Then we calculate the non-linear parameter of the
two racetrack inflation models.

\section{the non-linear parameter}

This section we summarize the result in \cite{a0510709}. For the
background, the effective action of the simple coupling system of
Einstein gravity and scalar fields with an arbitrary inflation
potential $V(\varphi)$ is
\begin{equation}
\label{action}S=\int{\sqrt{-g}d^4x[\frac{M_p^2}{2}R-\frac{1}{2}
G_{IJ}\partial_\mu\varphi^I\partial^\mu\varphi^J-V(\varphi)]},
\end{equation}
where $G_{IJ}\equiv G_{IJ}(\varphi)$ represents the metric on the
manifold parameterized by the scalar field values. And $8\pi
G=M_p^{-2}$ represents the reduced Planck mass. Units are chosen
such that $c=\hbar=1$. And the Friedmann-Robertson-Walker metric is
used,
\begin{equation}
\label{FRWmetric}ds^2=-dt^2+a^2(t)\delta_{ij}dx^idx^j.
\end{equation}

The non Gaussianity of the curvature is expressed in the form:
\begin{equation}
\label{defNonLinear}\zeta=\zeta_g-\frac{3}{5}f_{NL}(\zeta_g^2-\langle\zeta_g^2\rangle),
\end{equation}
where $\zeta_g$ is Gaussian, with $\langle\zeta_g\rangle=0$. And
$f_{NL}$ is the non-linear parameter.

On the other hand, the curvature perturbation can be expressed as
the difference between an initial space-flat fixed-$t$ slice and a
final uniform energy density fixed-$t$ slice (see
\cite{Starobinsky,a0504045,a9507001} for details),
\begin{equation}
\label{curvaturePer}\zeta(t,\textbf{x})=\delta N,
\end{equation}
where $N=\int{Hdt}=\int{\frac{\dot{a}}{a}dt}$ is the integrated
number of e-folds. Here and after, one dot denotes the derivative
with respect to the time, $\dot{a}=da/dt$. Expanding the curvature
perturbation to the second order \cite{a0504045}, we get
\begin{equation}
\label{deltaN}\zeta\simeq N_{,I}(t)\delta\varphi^I(\textbf{x})+
\frac{1}{2}N_{,IJ}(t)\delta\varphi^I(\textbf{x})\delta\varphi^J(\textbf{x}),
\end{equation}
where $N_{,I}=\frac{\partial N}{\partial\varphi^I}$,
$N_{,IJ}=\frac{\partial^2 N}{\partial\varphi^I\partial\varphi^J}$.

Equating the bispectrums of $\zeta$ obtained by using the two
equations (\ref{deltaN}) and (\ref{defNonLinear}), we may get
\begin{equation}
\label{FNL}f_{NL}=-\frac{5}{6}\times\frac{G^{IM}G^{KN}N_{,I}N_{,K}N_{,MN}}
{(N_{,I}N_{,J}G^{IJ})^2},
\end{equation}
where $G^{IJ}$ are the elements of the inverse metric of the field
space. In this result, the intrinsic non-Gaussian part,
$\sim\langle\delta\varphi^I(\textbf{k}_1)
\delta\varphi^J(\textbf{k}_2)\delta\varphi^K(\textbf{k}_3)\rangle$,
has been neglected.

Define one parameter, $\varepsilon_I$, as
\begin{equation}
\label{epsilonLIV}\varepsilon_I\equiv-\frac{V_{,I}M_p}{\sqrt{2}V},
\end{equation}
where $V_{,I}\equiv\frac{\partial V}{\partial\varphi^I}$. Then,
using the slow-rolling approximation, the slow-rolling parameter,
$\varepsilon$, can be expressed as
\begin{equation}
\label{epsilonV}\varepsilon\equiv-\frac{\dot{H}}{H^2}\simeq
G^{IJ}\varepsilon_I\varepsilon_J.
\end{equation}
Further, neglecting the non-adiabatic perturbations of the inflation
fields, we can get
\begin{equation}
\label{dNDPhi}N_{,I}\simeq-\frac{1}{2\varepsilon
V}V_{,I}=\frac{\varepsilon_I}{\sqrt{2}\varepsilon M_p}.
\end{equation}
Now, the non-linear parameter may be expressed as
\begin{equation}
\label{fNLSlowRollAppr}f_{NL}\simeq\frac{5M_p}{3\varepsilon}\times\beta,
\end{equation}
with
\begin{equation}
\label{beta}\beta=\frac{\varepsilon^2}{M_p}+
\frac{{G^{MN}}_{,L}}{\sqrt2}G^{JL}\varepsilon_M\varepsilon_N\varepsilon_J-
\frac{1}{2M_p}G^{JL}G^{MN}\varepsilon_J\varepsilon_M\eta_{NL},
\end{equation}
where ${G^{MN}}_{,L}\equiv\frac{\partial G^{MN}}{\partial\varphi^L}$
and $\eta_{NL}\equiv\frac{V_{,IJ}M_p}{V},\
V_{,IJ}\equiv\frac{\partial^2V}{\partial\varphi^I\partial\varphi^J}$.

\section{non-Gaussianity of the racetrack model with single modulus}

\label{singlemodulus} In \cite{t0406230}, basing on a simple
extension of \textbf{KKLT} scenario \cite{KKLT}, the authors suggest
a racetrack inflationary model in the frame of string theory. This
model is equivalent to a double-field inflationary model, with the
effective action
\begin{equation}
\label{RTaction}S=\int{\sqrt{-g}d^4x[\frac{1}{2}R-
\frac{3M_p^2}{4X^2}(\partial_\mu X\partial^{\mu}X+\partial_\mu
Y\partial^{\mu}Y)-V(X,Y)]},
\end{equation}
with
\begin{eqnarray}
V(X,Y)&=&\frac{E}{X^\alpha}
+\frac{e^{-aX}}{6X^2}[aA^2(aX+3)e^{-aX}+3W_0aA\cos(aY)] \nonumber \\
&+&\frac{e^{-bX}}{6X^2}[bB^2(bX+3)e^{-bX}+3W_0bB\cos(bY)] \nonumber\\
&+&\frac{e^{-(a+b)X}}{6X^2}[AB(2abX+3a+3b)\cos((a-b)Y)].\label{RTPotential}
\end{eqnarray}
In this model, $X$ and $Y$ correspond to the scalar fields in
Eq.(\ref{action}), $\varphi^I,\ I=1,2$. $E$, $\alpha$, $a$, $A$,
$b$, $B$ and $W_0$ are constant parameters of this model. The
non-zero components of the field-space metric are
$G_{11}=G_{22}=\frac{3M_p^2}{2X^2}$. In \cite{t0406230}, the
appropriate values of the parameters have been suggested,
\begin{eqnarray}
E&=&4.14668\times10^{-12},\ \alpha=2,\nonumber\\
a&=&\frac{2\pi}{100},\ A=\frac{1}{50}, \ b=\frac{2\pi}{90},\
B=-\frac{35}{1000},\ W_0=-\frac{1}{5000},\label{parametersVal}
\end{eqnarray}
The inflationary saddle point is at
\begin{equation}
X_{saddle}=123.22,\ \ \ \ \ \ Y_{saddle}=0.
\end{equation}
The units of the values above have been taken to be $M_p=1$.

Setting the initial point of the inflation at
\[
X=123.22, \ Y=0.2,
\]
and then calculating the slow-rolling parameters at the COBE scale
or equivalently at $N\simeq60$ which is the number of e-foldings
before the end of the inflation, we get
\begin{equation}
\label{fNLNumericalVal}f_{NL}\simeq-0.16.
\end{equation}
The observation \cite{WMAP3} shows that the limits on primordial
non-Gaussianity are $-54<f_{NL}<114$ at the $95\%$ confidence level.
So this result fits the limits. Supposing the contribution of the
non-adiabatic part and the intrinsic non-Gaussianity at the same
order of this result, we may expect that the total non-Gaussianity
of this model should be $|f_{NL}|\sim 1$. This is still a reasonable
result.

\section{non-Gaussianity of the racetrack model with two K\"{a}hler moduli}

\label{twoemoduli} In \cite{t0603129}, by compactifying to the
orientifold of degree 18 hypersurface $\mathbb{P}^4_{[1,1,1,6,9]}$,
an elliptic fibered Calabi-Yau over $\mathbb{P}^2$, the authors
suggest a racetrack inflation model with tow K\"{a}hler moduli. The
effective action of this models is
\begin{equation}
\label{RT2action}S=\int{\sqrt{-g}d^4x[\frac{1}{2}R-\mathcal
{L}_{kin}-V]},
\end{equation}
with
\begin{eqnarray}
\mathcal {L}_{kin}&=&\frac{3}{8(X^{3/2}_1-X^{3/2}_2)^2}\times
\{\frac{2X^{3/2}_1+X^{3/2}_2}{\sqrt{X_1}}(\partial X^2_1+\partial
Y^2_1) \nonumber \\
&-&6\sqrt{X_1X_2}(\partial X_1\partial X_2+\partial Y_1\partial Y_2) \nonumber \\
&+&\frac{X^{3/2}_1+2X^{3/2}_2}{\sqrt{X_2}}(\partial X^2_2+\partial
Y^2_2)\},\label{RT2KineticTerm}
\end{eqnarray}
and
\begin{eqnarray}
V&=&\frac{D}{({X_2}^{3/2}-{X_1}^{3/2})^2}+\frac{216}{({X_2}^{3/2}-{X_1}^{3/2})^2}\times
\nonumber \\
&\{&B^2b(b{X_2}^2+2b{X_1}^{3/2}{X_2}^{1/2}+3X_2)e^{-2bX_2} \nonumber \\
&+&A^2a(3X_1+2a{X_2}^{3/2}{X_1}^{1/2}+a{X_1}^2)e^{-2aX_1} \nonumber\\
&+&3BbW_0X_2e^{-bX_2}\text{cos}(bY_2)+3AaW_0X_1e^{-aX_1}\text{cos}(aY_1) \nonumber \\
&+&3ABe^{-aX_1-bX_2}(aX_1+bX_2+2abX_1X_2)\text{cos}(-aY_1+bY_2)\}.\label{RT2Potential}
\end{eqnarray}
In this model, $X_1,\ Y_1$ and $X_2,\ Y_2$ correspond to the scalar
fields in Eq.(\ref{action}), $\varphi^I$, with $I=1,2,3,4$. And $D,\
W_0,\ a,\ A,\ b,\ B$ are the constant parameters of this model. In
\cite{t0603129}, the appropriate values of these parameters are
suggested,
\begin{eqnarray}
D&=&6.21\times10^{-9},\ \ W_0=5.227\times10^{-6}, \\
A&=&0.56,\ \ B=7.47\times10^{-5},\ \ a=\frac{2\pi}{40},\ \
b=\frac{2\pi}{258}.
\end{eqnarray}
And the inflationary saddle point is at
\begin{equation}
X_1=108.96,\ \ X_2=217.69,\ \ Y_1=20,\ \ Y_2=129.
\end{equation}
As before, the units of the values above are taken to be $M_p=1$.

In \cite{t0603129}, an example of the inflation with 980 e-foldings
is suggested. Again, Calculating the slow-rolling parameters at the
COBE scale, we get
\begin{equation}
\label{fNL2NumericalVal}f_{NL}\simeq 0.02.
\end{equation}
Here, we note that we have taken the COBE scale at $N\simeq60$ as in
the section \ref{singlemodulus} in order to compare between the both
models. This is also a reasonable result.

\section{Summary and Discussion}

Above, we have obtained the non-Gaussianities of the models in
\cite{t0406230,t0603129} by using the result in \cite{a0510709}.
Both of the models give the reasonable results, although the two
results are different. By considering that in our result, the
contribution of the intrinsic non-Gaussianity of the fields and the
non-adiabatic perturbations are neglected, we do not think the
difference is important.

However, our results illustrate that there exits remarkable
difference between multi-field inflation models with the nontrivial
metric of the field space and multi-field inflation models with the
trivial metric of the field space (or single field inflation
models). In \cite{a0506056}, it has been revealed that, for a
multi-field model with $G_{IJ}=\delta_{IJ}$, the contribution of
Eq.(\ref{FNL}) is about at the order of $\varepsilon$. But this
analysis can not be applied to the models in
\cite{t0406230,t0603129}. In fact, for the model in \cite{t0406230},
at the COBE scale, we get
\[
\varepsilon\simeq3\times10^{-10}.
\]
And for the model in \cite{t0603129}, at the COBE scale, we get
\[
\varepsilon\simeq9\times10^{-10}.
\]
So the non-Gaussianities of the two models are very large  compared
with the values of $\varepsilon$.

This is due to the third term on the right-hand side of
Eq.(\ref{beta}). For the first two terms on the right-hand side of
Eq.(\ref{beta}), they are expected to be at the order of
$\varepsilon^2$. But the third term,
\[
G^{JL}G^{MN}\varepsilon_J\varepsilon_M\eta_{NL},
\]
is different. We may expect that this term be at the order of
$\varepsilon G^{MN}\eta_{NL}$, basing on the relation,
$\varepsilon=G^{IJ}\varepsilon_I\varepsilon_J$. However, for a
multi-field inflation with a non-trivial metric of field space, some
components of the metric, $G^{MN}$, may be large. Then, even if
$\eta_{IJ}$ is at the order of $\varepsilon$, the contribution of
this term would be much larger than the first two terms.

This is just the case for the models in \cite{t0406230,t0603129}. In
the model of \cite{t0406230}, $G^{MN}$ is about $10^4$ and
$\eta_{IJ}$ is about $10^{-4}$. Considering the prefactors, we
expect the result (\ref{fNLNumericalVal}) as obtained. For the model
of \cite{t0603129}, $G^{MN}$ is about $10^4$ and $\eta_{IJ}$ is
about $10^{-2}$. Then we may expect the non-Gaussianity be at the
order $10^2$. However, this is not the case because the opposite
contribution of some components reduces the value. So we just obtain
the result (\ref{fNL2NumericalVal}). But this reduction does not
work during the initial stage of the inflation with 980 e-foldings
in \cite{t0603129}. Calculating the non-Gaussianity at the initial
point of the inflation, we get $f_{NL}\simeq-125$. This value is
beyond the limits. Fortunately, for a inflation with 980 e-foldings,
the scales of the perturbations generated during the initial stage
of inflation are so large that the large non-Gaussianity calculated
at the initial point does not matter the observation. However, our
analysis above reveals that the large non-Gaussianity may be
generated in a multi-field inflation model.

\section*{Acknowledgments}
We acknowledge the useful discussions with J. Cline, A. Linde, R.
Kallosh, C. Burgess and G.R. Marta.

\end{document}